\documentclass[aps,prb,twocolumn,showpacs,superscriptaddress]{revtex4-1}

\usepackage[pdftex]{graphicx} 
\usepackage{natbib}
\usepackage{booktabs}
\newcommand{\bv}[1]{\mathbf{#1}}
\newcommand{\bvv}[1]{{\bf#1}}

\newcommand{\one}{Fig.~\ref{f1}}
\newcommand{\two}{Fig.~\ref{f2}}
\newcommand{\three}{Fig.~\ref{f3}}

\newcommand{\etal}{\textit{et al.}}

\newcommand{\dxy}{$d_{xy}$}
\newcommand{\dxz}{$d_{xz/yz}$}

\newcommand{\Py}{$P_{\textrm{\small{y}}}$}

\newcommand{\IiL}{$I_{i}^{\textrm{\tiny{L}}}$}
\newcommand{\IiR}{$I_{i}^{\textrm{\tiny{R}}}$}

\newcommand{\Iiup}{$I_{i}^{\uparrow}$}
\newcommand{\Iidown}{$I_{i}^{\downarrow}$}
\newcommand{\Iiud}{$I_{i}^{\uparrow \downarrow}$}

\begin{document}
\title{Absence of Giant Spin Splitting in the Two-Dimensional Electron Liquid at the Surface of SrTiO$_3$ (001)}

\author{S. McKeown Walker}
\affiliation{Department of Quantum Matter Physics, University of Geneva, 24 Quai Ernest-Ansermet, 1211 Geneva 4, Switzerland}
\author{S. Ricc\`o}
\affiliation{Department of Quantum Matter Physics, University of Geneva, 24 Quai Ernest-Ansermet, 1211 Geneva 4, Switzerland}
\author{F. Y. Bruno}
\affiliation{Department of Quantum Matter Physics, University of Geneva, 24 Quai Ernest-Ansermet, 1211 Geneva 4, Switzerland}
\author{A. de la Torre}
\affiliation{Department of Quantum Matter Physics, University of Geneva, 24 Quai Ernest-Ansermet, 1211 Geneva 4, Switzerland}
\author{A. Tamai}
\affiliation{Department of Quantum Matter Physics, University of Geneva, 24 Quai Ernest-Ansermet, 1211 Geneva 4, Switzerland}
\author{E. Golias}
\affiliation{Helmholtz-Zentrum Berlin f\"ur Materialien und Energie, Albert-Einstein-Str. 15, 12489 Berlin, Germany}
\author{A. Varykhalov}
\affiliation{Helmholtz-Zentrum Berlin f\"ur Materialien und Energie, Albert-Einstein-Str. 15, 12489 Berlin, Germany}
\author{D. Marchenko}
\affiliation{Helmholtz-Zentrum Berlin f\"ur Materialien und Energie, Albert-Einstein-Str. 15, 12489 Berlin, Germany}
\author{M. Hoesch}
\affiliation{Diamond Light Source, Harwell Campus, Didcot, United Kingdom}
\author{M.S. Bahramy}
\affiliation{Quantum-Phase Electronics Center, Department of Applied Physics, The University of Tokyo, Tokyo 113-8656, Japan}
\affiliation{RIKEN Center for Emergent Matter Science (CEMS), Wako 351-0198, Japan}
\author{P.D.C. King}
\affiliation{SUPA, School of Physics and Astronomy, University of St Andrews, St Andrews, Fife KY16 9SS, United Kingdom}
\author{J. S\'anchez-Barriga}
\affiliation{Helmholtz-Zentrum Berlin f\"ur Materialien und Energie, Albert-Einstein-Str. 15, 12489 Berlin, Germany}
\author{F. Baumberger}
\affiliation{Department of Quantum Matter Physics, University of Geneva, 24 Quai Ernest-Ansermet, 1211 Geneva 4, Switzerland}
\affiliation{Swiss Light Source, Paul Scherrer Institut, CH-5232 Villigen PSI, Switzerland}

\date{\today}

\begin{abstract}
{We reinvestigate the putative giant spin splitting at the surface of SrTiO$_3$ reported by Santander-Syro \textit{et al.} [Nature Mat. 13, 1085 (2014)]. Our spin- and angle-resolved photoemission experiments on (001) oriented surfaces supporting a two-dimensional electron liquid with high carrier density show no detectable spin polarization in the photocurrent. We demonstrate that this result excludes a giant spin splitting while it is fully consistent with the unconventional Rashba-like splitting seen in band structure calculations that reproduce the experimentally observed ladder of quantum confined subbands.}
\end{abstract}

\maketitle
Two-dimensional electron liquids (2DELs) formed at the interfaces between insulating transition metal oxides are important for the rapidly growing field of oxide electronics. Their potential utility lies in their exotic responses to external fields which, for the prototypical case of the LaAlO$_3$/SrTiO$_3$ (LAO/STO) interface, includes gate-tunable superconductivity \cite{Reyren2007,Caviglia2008} possibly coexisting with magnetism \cite{Li2011}, and gate-tunable Rashba interaction \cite{Caviglia2010,Fete2014,Hurand2015}. It has been shown that STO can support such a two-dimensional electron liquid in many other configurations, for example when interfaced with amorphous LAO~\cite{Liu2013}, by electrolyte gating \cite{Ueno2008} or by reduction of the bare surface by UV radiation~\cite{Meevasana2011a,McKeownWalker2015} or Al capping~\cite{Rodel2016}. Irrespective of their origin, all these systems show a similar electronic structure with multiple subbands and a characteristic orbital polarization, commonly understood as a consequence of quantum confinement of the Ti $t_{2g}$ states in a potential well induced by band bending~\cite{Stengel2011,Meevasana2011a,Santander-Syro2011a,Joshua2012,King2014,McKeownWalker2015,Salluzzo2009,Cancellieri2014}.

\em Ab initio\em~ density functional theory (DFT) of both interface and surface geometries predicts an unconventional Rashba-like spin splitting of these quantum confined subbands due to broken inversion symmetry and the interplay of orbital and spin degrees of freedom. The lifting of spin degeneracy is found to be of the order of~$\sim1$~meV at the Fermi surface except in the vicinity of avoided crossings of subbands with different orbital character where it can be enhanced by almost an order of magnitude~\cite{Zhong2013b,Khalsa2013a,King2014,Kim2014a,Kim2013a,Altmeyer2015,Garcia-Castro2015b}. The resulting $k$-space spin texture is complex and has not yet been observed experimentally. However, the magnitude and carrier density dependence of the Rashba splitting inferred from transport and quantum oscillation experiments \cite{Caviglia2010,Fete2014,Nakamura2012,Liang2015} is in good agreement with these calculations.

Recently, a completely different interpretation of the basic electronic structure of the 2DEL at the (001) surface of STO has been proposed by Santander-Syro \etal{} to explain a large spin polarization signal in their spin- and angle- resolved photoemission spectroscopy (SARPES) measurements\cite{Santander-Syro2014}. The authors of Ref.~\onlinecite{Santander-Syro2014} propose that the first two light subbands of the STO 2DEL (SB1, SB2 in \one) arise from a single band with a giant Rashba splitting of approximately 100 meV at the chemical potential. In order to reconcile this claim with the large subband splitting at the $\Gamma$ point that is well established from high-resolution angle-resolved photoemission spectroscopy (HR-ARPES)\cite{Santander-Syro2011a,Meevasana2011a,King2014,McKeownWalker2015,Rodel2016}, Santander-Syro \etal{} propose the existence of strong ferromagnetism with significant out-of-plane moments. To date, a Rashba splitting of this magnitude has not been reproduced experimentally or explained theoretically~\cite{Altmeyer2015,Ghosh2015,Garcia-Castro2015b}. Moreover, a giant Rashba splitting is inconsistent with transport measurements of both surface \cite{Nakamura2012,Lee2011c} and interface\cite{Caviglia2010,Fete2014} 2DELs in STO. It is also far greater than experimentally observed spin-splittings in other systems with broken inversion symmetry whose constituent atoms, like STO, have relatively low atomic numbers leading to weak atomic spin-orbit interaction\cite{Tamai2013b,Marchenko2013}.

Here we present low temperature SARPES meaurements on fractured STO that show a negligible spin polarization of the photocurrent. We demonstrate that this result is fully consistent with band structure calculations that reproduce the experimentally observed ladder of subbands as well as the Rashba splitting deduced from transport experiments, while it is inconsistent with the giant spin-splitting reported by Santander-Syro \etal\cite{Santander-Syro2014}.

\begin{figure*}[th!]
	\includegraphics[width=17.8cm]{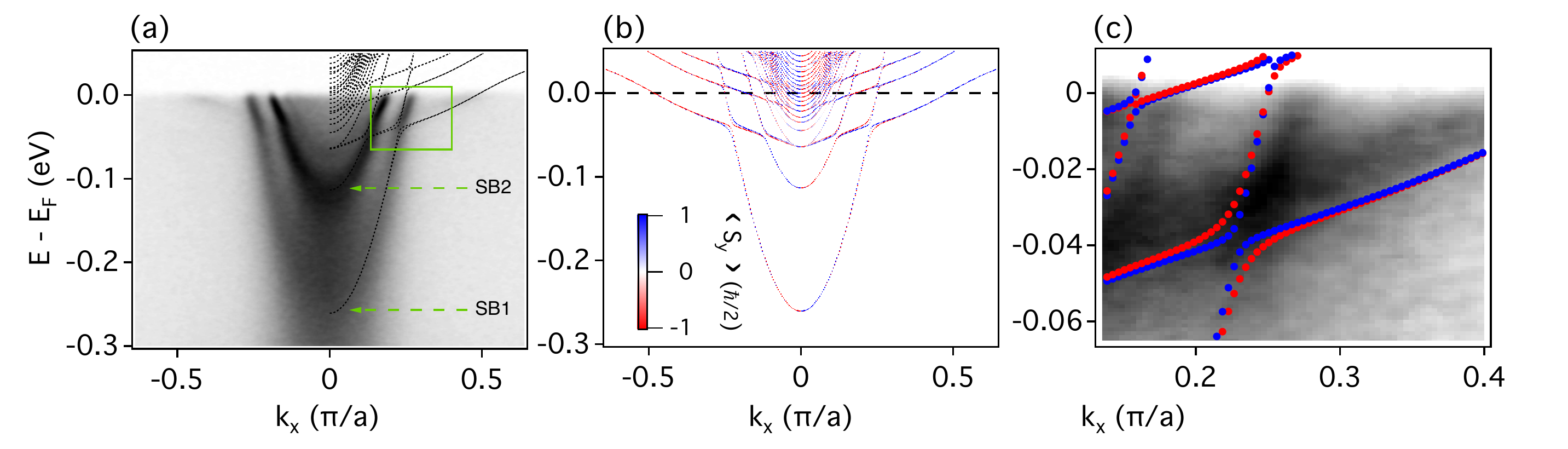}
	\caption{(color online)(a) Subband structure of the STO (001) surface 2DEL from HR-ARPES (greyscale plot) taken at 47 eV with $s$-polarized light along the [100] direction. The result of self-consistent tight binding supercell calculations is over-plotted (black dashed lines).(b) Band structure calculation from (a) with color-coded spin expectation values $\langle S_{y} \rangle$.(c) Region of energy-momentum space indicated by the green box in (a), measured with $p$-polarized light at 47 eV to enhance the intensity of the heavy subband which has out-of-plane orbital character.}
	\label{f1}
\end{figure*}

Single crystals of commercially grown (Crystal Base), lightly electron doped Sr$_{1-  x}$La$_{ x}$TiO$_{3}$(001) (\( x  = 0.001\)) were measured. The La doping results in a small residual bulk conductivity which  helps to eliminate charging effects during ARPES measurements but does not otherwise influence our results. Data were taken using photon energies in the range 45 -- 100 eV. HR-ARPES measurements were performed at a temperature of 10 K with a Scienta R4000 hemispherical analyser at the I05 beamline of the Diamond Light Source with angular resolution $< 0.3^\circ$ and energy resolution $<15$ meV and pressures $<{1 \times 10 ^{-10}}$ mbar. SARPES experiments were performed at a temperature of 20 K at pressures $< 1\times10^{-10}$ mbar, using polarized undulator radiation at the UE112-PGM1 beamline of BESSY II. Spin analysis of the photoelectrons was provided by a Rice University Mott-type spin polarimeter \cite{Burnett-RSI-94} operated at 26 kV and coupled to a SPECS Phoibos 150 hemispherical analyzer. The energy resolution of the SARPES experiment was $\sim$100 meV, the angular resolution $\sim$0.8$^{\circ}$, and the Shermann function $S_{\rm{eff}}=0.16$. Samples were fractured \textit{in situ} at the measurement temperature and pressure. 

 \begin{figure*}[th!]
 	\includegraphics[width=17.8cm]{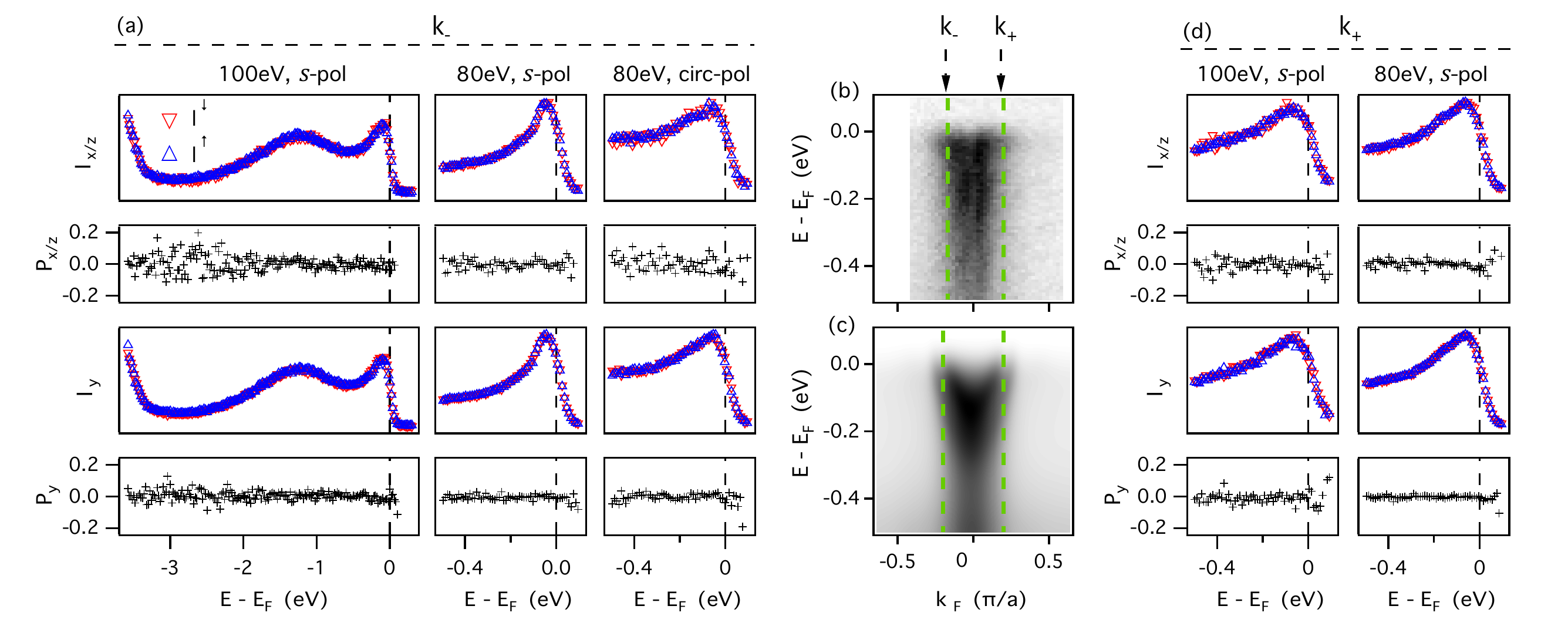}
 	\caption{(color online)(a) Spin-resolved photoemission from the STO (001) surface 2DEL at $k_{-}$ defined in (b). Photon energy and polarization are indicated for each column. First row panels: Spin-up (blue symbols) and spin-down (red symbols) energy distribution curves for the x/z component of the spin polarization vector calculated from Eq. \ref{Iupdown}. Second row panel: the corresponding polarization calculated from Eq.~\ref{P} with $S_{\rm{eff}}=0.16$. Third and fourth row panels, as first and second row for the y component of the spin polarization vector. (b) Spin integrated dispersion taken with 80~eV $s$-polarized photons. (c) High resolution spin-integrated dispersion from \one(a) convolved with a 2D Gaussian of width 0.06 \AA$^-1$ and 90 meV to simulate the experimental resolution. (d) As (a) at the momentum $k_{+}$ as defined in (b).}
 	\label{f2}
 \end{figure*}
 
We first demonstrate that our HR-ARPES data is fully consistent with a subband structure resulting from quantum confinement of the STO conduction band and the ensuing unconventional Rashba spin splitting predicted by several authors \cite{Zhong2013b,Khalsa2013a,King2014,Kim2014a,Kim2013a,Garcia-Castro2015b,Altmeyer2015}. \one~(a) shows the energy-momentum dispersion of the (001) STO surface 2DEL measured with spin-integrated HR-ARPES. Three bands with a light band mass and a fourth with a comparatively heavy band mass can be identified. The light bands have \dxy~orbital character and are more spatially confined than the heavy band of \dxz~orbital character~\cite{King2014}. The ordering and confinement energies of these subbands are in good agreement with the band structure calculation following the approach of Refs.~\onlinecite{Bahramy2012,King2014} that is overlaid on the right hand side of \one(a).
This calculation is the self-consistent solution of coupled Poisson-Schr\"odinger equations with a tight-binding supercell Hamiltonian obtained from transfer integrals generated by downfolding ~\em ab initio\em~ DFT wave functions onto maximally localized Wannier functions.
A band bending potential has been included as an on-site potential term. The electrostatic boundary conditions are chosen to conserve bulk charge neutrality and reproduce the total experimental bandwidth of $\approx 250$ meV. 
We include an electric field dependent dielectric constant of the form suggested in Ref.~\onlinecite{Copie2009}.
 Further details of the calculation can be found elsewhere\cite{Bahramy2012,King2014}.

A small Rashba-like spin splitting is apparent throughout the calculated subband structure. The spin expectation value  $\langle S_{y} \rangle$ of each eigenstate is represented by the red-white-blue color scale in \one(b) and corresponds to spins locked perpendicular to the momentum. This is the characteristic signature of the Rashba interaction resulting from broken inversion symmetry, which in this case arises from the band bending potential at the surface. However, it is evident from \one~(c) that the spin splitting deviates from a conventional Rashba picture near avoided crossings of light and heavy bands. 
In these limited regions of $k$-space the wave functions are linear combinations of the $t_{2g}$ crystal field eigenstates. Hence, their orbital angular momentum \bvv{L} is no longer fully quenched leading to a sizable spin-orbit coupling $\bv{L}\cdot \bv{S}$ and thus an enhanced spin splitting~\cite{King2014}. A more detailed comparison to the data shown in \one(c) highlights that the predicted spin spitting, even at the avoided crossings where it can be as large at 7 meV, would be obscured by resolution and lifetime broadening in HR-ARPES, and as such is consistent with the experimentally determined subband structure. 

While the good agreement between our band structure calculations and spin-integrated HR-ARPES data is hard to reconcile with a giant Rashba splitting as reported by Ref.~\onlinecite{Santander-Syro2014}, it does not fully exclude it.
Therefore, to reinvestigate this discrepancy, we performed new SARPES measurements on fractured surfaces of (001) orientated La doped STO, as were used in \one{}. To characterize the sample surface prior to the spin-resolved measurements, we acquired the spin-integrated dispersion shown in \two(b), which confirms the presence of a 2DEL. The lower data quality as compared to \one{} can be attributed to resolution effects. To demonstrate this we show in \two~(c) a two-dimensional convolution of the HR-ARPES measurement in \one(a) with a Gaussian, representing the energy and momentum resolution of our SARPES measurements. Comparison of \two(b) and \two(c) confirms that the states have similar carrier density. The slight differences in spectral weight distribution between \two~(b) and (c) can be attributed to the photon energy dependence of matrix elements, which increases the contribution of the shallow heavy subband in \two~(b).

Our SARPES measurements are sensitive to the two components of the spin polarization vector in the Mott scattering plane. In the sample reference frame these correspond to the $y$ component of the spin polarization vector that lies entirely in the surface plane and is perpendicular to the electron momentum $k_x$, and a combination of the $x$-component and out-of-plane $z$-component. The corresponding spin-resolved energy distribution curves (EDCs) are denoted by \Iiud~where $i=y$ or $i=x/z$ respectively. These are calculated from the left and right channeltron count rates \IiL~and \IiR~using the standard expression
\begin{equation}
I_i^{\uparrow \downarrow}=\frac{1}{2} (I_{i}^L + I_{i}^R) (1 \pm P_i)
\label{Iupdown}
\end{equation}
where $P_i$ is the spin polarization given by
\begin{equation}
P_i=\frac{1}{S_{\rm{eff}}} \frac{I_{i}^{L}-I_{i}^{R}}{I_{i}^{L}+I_{i}^{R}}=\frac{I_{i}^{\uparrow}-I_{i}^{\downarrow}}{I_{i}^{\uparrow}+I_{i}^{\downarrow}}
\label{P}
\end{equation}
and $S_{\rm{eff}}=0.16$ is the effective Sherman function. Prior to the calculation of \Iiud~we subtracted a constant background from the EDCs \IiL~and \IiR, to account for detector dark counts and the photocurrent due to higher harmonics of the exciting radiation. Subsequently each EDC pair has been normalized in an energy window where no spin polarization is expected (such as the valence band maximum) to account for different detector sensitivities.

In \two(a) and (d) the spin-resolved spectra \Iiup~(blue symbols) and \Iidown~(red symbols) at the momenta $k_\pm$ indicated in \two(b) are shown together with their corresponding polarization signal $P_i$ (black symbols). It is evident from this data that all components of the spin-polarization measured at different photon energies and polarizations are below the noise level. In particular, we do not see any signatures of a Rashba-like spin-splitting, which would be expected in the y channel within $\approx0.3$~eV of the Fermi level for the STO surface 2DEL.
The upper limit on polarization features that may be obscured by noise in our measurements is $\sim0.05$, which is far smaller than the spin-polarization reported in Ref.~\onlinecite{Santander-Syro2014}. 
As we will show in the following, our measurements rule out a giant Rashba splitting, while they are fully consistent with the much smaller, unconventional Rashba splitting found consistently in our band structure calculations and by several other authors~\cite{Zhong2013b,Khalsa2013a,King2014,Kim2014a,Kim2013a,Altmeyer2015,Garcia-Castro2015b}.

\begin{figure*}[th!]
	\includegraphics[width=17.8cm]{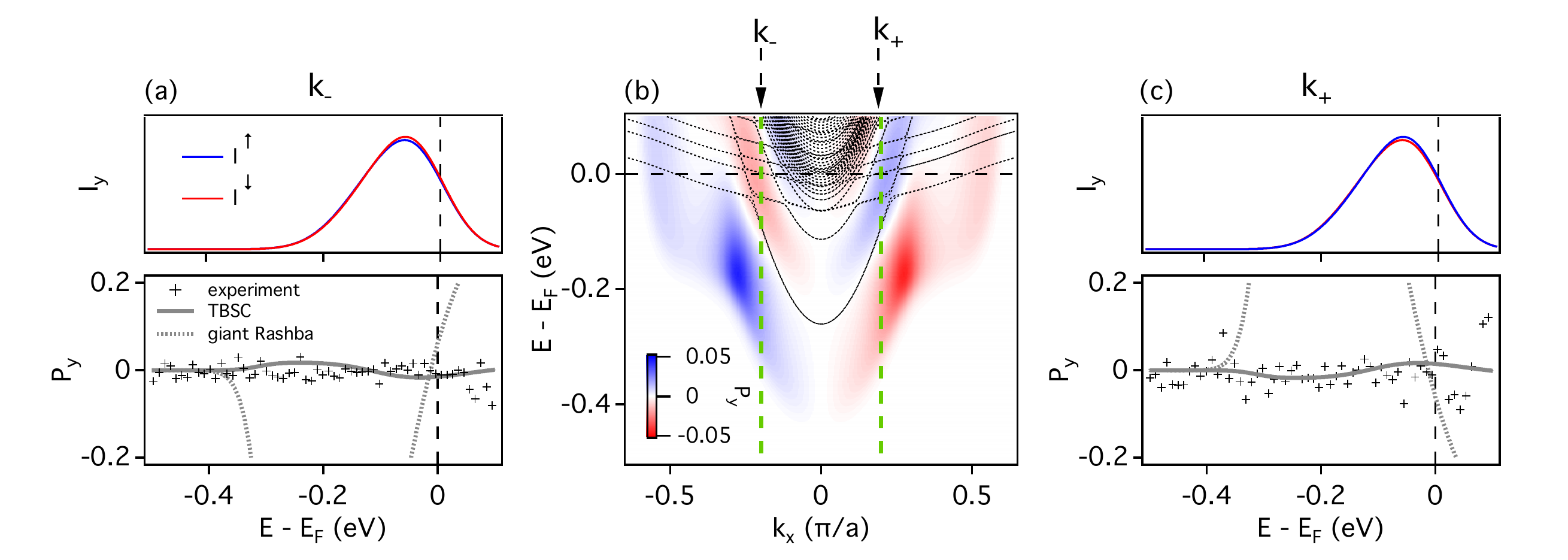}
	\caption{(color online) (a) Upper panel: Simulated spin-resolved photocurrent intensity $I_{\rm{y}}^{\uparrow \downarrow}$ (blue and red solid lines) at the momentum $k_-$ defined in (b) for the subband structure resulting form tight binding supercell (TBSC) calculations. Lower panel: The polarization signal \Py~corresponding to $I_{\rm{y}}^{\uparrow \downarrow}$ above (solid grey lines). The simulated \Py~signal for a single parabolic band dominated by a 100 meV Rashba spin splitting and Zeeman-like degeneracy lifting at $k_x=0$ (dashed grey lines). Data from \two~are over-plotted (black cross symbols).(b) The full energy and momentum dispersion of the simulated spin polarization signal $P_y$ for the TBSC subband structure (black dashed lines) is represented by the red-white-blue colour scale. (c) As (a), for the momentum $k_+$ defined in (b).}
	\label{f3}
\end{figure*}

In order to facilitate the discussion of our SARPES results, in \three{} we show a minimal simulation of the spin-polarization of the photocurrent expected from the fully spin-polarized initial states of our band structure calculations shown in \one. To this end we adopt a non-interacting single particle description of the photoemission process and neglect all matrix element effects. In this simple model,
$I^{\uparrow \downarrow}_{\rm{y}}$ are found by weighting the poles of the spectral function by the probability $\frac{1}{2} \pm \frac{1}{\hbar} \langle S_y \rangle$. Experimental conditions are taken into account by multiplying $I^{\uparrow \downarrow}_{\rm{y}}$ with a Fermi function at the measurement temperature, and convolving the spectral function with a 2D Gaussian of width 120~meV and 0.08~\AA$^{-1}$ (corresponding to 0.8$^{\circ}$). EDCs for this simulated $I^{\uparrow \downarrow}_{\rm{y}}$ are shown in \three(a) and (c) (blue and red lines) for the same $k_\pm$ as were measured experimentally. The effect of the experimental resolution is evident in the presence of a single broad peak instead of multiple sharp peaks at the energy-momentum positions of the eigenvalues of the calculation in \one. This broad peak nevertheless shows a small asymmetry which changes sign with $k$. 
The \Py{} corresponding to these EDCs is found using Eq.~\ref{P} and is plotted in \three{}(a) and (c) as solid grey lines. The features of this simulated polarization signal are $<0.02$, which is below the noise level in our data (black cross symbols). Hence, our experimental resolution, which is similar to the one in Ref.~\onlinecite{Santander-Syro2014}, completely masks the spin polarization of the initial state. The absence of a significant polarization signal in our SARPES data is thus fully consistent with the spin-polarized subband structure shown in \one.

We further note that the simulated polarization signal, shown in \three~(b) over the full range of the 2DEL dispersion, has little similarity with the corresponding initial state polarization shown in \one~(b). SARPES data from complex systems such as the STO 2DEL are thus highly prone to misinterpretation. Indeed, it is nearly impossible to deconvolve the simulation and unambiguously deduce the initial state polarization from the \Py{} signal shown in \three~(b).
Even with much improved resolution and detailed knowledge of the subband structure, it would remain challenging to extract the spin texture following the usual fitting procedures.
Moreover, in cases like the STO 2DEL, where the predicted spin-splitting is smaller than the intrinsic quasiparticle lifetime broadening over most of energy-momentum space, spin-interference effects will further complicate the interpretation of SARPES data~\cite{Meier2011}. In this case the initial state spin polarization can no longer be understood from band structure calculations that neglect interactions and the entire notion of well defined spin-states becomes questionable. 

Next, we use the same approach as above to simulate \Py{} for the single band electronic structure with a giant Rashba splitting of 100~meV and a Zeeman gap at the $\Gamma$ point as proposed by Santander-Syro \etal~\cite{Santander-Syro2014}.
This results in a polarization up to $P_{y}= \pm0.9$, shown as grey dashed lines in \three(a) and (c), which is significantly greater than the noise level in our measurements. We can thus unambiguously exclude a Rashba splitting of this magnitude in our data.

In conclusion, we have presented SARPES data from the STO (001) surface 2DEL prepared on \em in-situ\em{} fractured surfaces with a band dispersion in excellent agreement with all published HR-ARPES data~\cite{Meevasana2011a,Santander-Syro2011a,Plumb2014,King2014,McKeownWalker2015,Rodel2016,Wang2015}. Our SARPES measurements do not show any significant spin-polarization signal which is consistent with the predicted small, unconventional Rashba splitting of the 2DEL subbands. These results exclude the possibility of a giant spin splitting as it was reported in Ref.~\onlinecite{Santander-Syro2014}. 
The origin of this discrepancy remains unclear. 
It could be related to differences in the surfaces measured in each case, however this seems unlikely given that the subband structure observed by conventional ARPES is ubiquitous and observed consistently for very different surface preparations~\cite{Meevasana2011a,Santander-Syro2011a,Plumb2014,King2014,McKeownWalker2015,Rodel2016,Wang2015}.
Alternatively, Altmeyer \etal~\cite{Altmeyer2015} proposed that an SARPES signal as was observed in Ref.~\onlinecite{Santander-Syro2014} might arise from ferromagnetic domains with exchange-split bands and a remnant Rashba-like spin texture. However, the small in-plane spin component calculated for this scenario is difficult to rationalize with the large spin polarization signal found in Ref.~\onlinecite{Santander-Syro2014}.

This work was supported by the Swiss National Science Foundation (Grant No. 200021-146995). We acknowledge Diamond Light Source for time on beamline I05 under Proposal No. SI11741. Financial support from the Deutsche Forschungsgemeinschaft (Grant No. SPP 1666) and the Impuls-und Vernetzungsfonds der Helmholtz-Gemeinschaft (Grant No. HRJRG-408) is gratefully acknowledged.
\bibliography{STO_SARPES.bib}

\begin{thebibliography}{36}%
\makeatletter
\providecommand \@ifxundefined [1]{%
 \@ifx{#1\undefined}
}%
\providecommand \@ifnum [1]{%
 \ifnum #1\expandafter \@firstoftwo
 \else \expandafter \@secondoftwo
 \fi
}%
\providecommand \@ifx [1]{%
 \ifx #1\expandafter \@firstoftwo
 \else \expandafter \@secondoftwo
 \fi
}%
\providecommand \natexlab [1]{#1}%
\providecommand \enquote  [1]{``#1''}%
\providecommand \bibnamefont  [1]{#1}%
\providecommand \bibfnamefont [1]{#1}%
\providecommand \citenamefont [1]{#1}%
\providecommand \href@noop [0]{\@secondoftwo}%
\providecommand \href [0]{\begingroup \@sanitize@url \@href}%
\providecommand \@href[1]{\@@startlink{#1}\@@href}%
\providecommand \@@href[1]{\endgroup#1\@@endlink}%
\providecommand \@sanitize@url [0]{\catcode `\\12\catcode `\$12\catcode
  `\&12\catcode `\#12\catcode `\^12\catcode `\_12\catcode `\%12\relax}%
\providecommand \@@startlink[1]{}%
\providecommand \@@endlink[0]{}%
\providecommand \url  [0]{\begingroup\@sanitize@url \@url }%
\providecommand \@url [1]{\endgroup\@href {#1}{\urlprefix }}%
\providecommand \urlprefix  [0]{URL }%
\providecommand \Eprint [0]{\href }%
\providecommand \doibase [0]{http://dx.doi.org/}%
\providecommand \selectlanguage [0]{\@gobble}%
\providecommand \bibinfo  [0]{\@secondoftwo}%
\providecommand \bibfield  [0]{\@secondoftwo}%
\providecommand \translation [1]{[#1]}%
\providecommand \BibitemOpen [0]{}%
\providecommand \bibitemStop [0]{}%
\providecommand \bibitemNoStop [0]{.\EOS\space}%
\providecommand \EOS [0]{\spacefactor3000\relax}%
\providecommand \BibitemShut  [1]{\csname bibitem#1\endcsname}%
\let\auto@bib@innerbib\@empty
\bibitem [{\citenamefont {Reyren}\ \emph {et~al.}(2007)\citenamefont {Reyren},
  \citenamefont {Thiel}, \citenamefont {Caviglia}, \citenamefont {Kourkoutis},
  \citenamefont {Hammerl}, \citenamefont {Richter}, \citenamefont {Schneider},
  \citenamefont {Kopp}, \citenamefont {R{\"{u}}etschi}, \citenamefont
  {Jaccard}, \citenamefont {Gabay}, \citenamefont {Muller}, \citenamefont
  {Triscone},\ and\ \citenamefont {Mannhart}}]{Reyren2007}%
  \BibitemOpen
  \bibfield  {author} {\bibinfo {author} {\bibfnamefont {N.}~\bibnamefont
  {Reyren}}, \bibinfo {author} {\bibfnamefont {S.}~\bibnamefont {Thiel}},
  \bibinfo {author} {\bibfnamefont {A.~D.}\ \bibnamefont {Caviglia}}, \bibinfo
  {author} {\bibfnamefont {L.~F.}\ \bibnamefont {Kourkoutis}}, \bibinfo
  {author} {\bibfnamefont {G.}~\bibnamefont {Hammerl}}, \bibinfo {author}
  {\bibfnamefont {C.}~\bibnamefont {Richter}}, \bibinfo {author} {\bibfnamefont
  {C.~W.}\ \bibnamefont {Schneider}}, \bibinfo {author} {\bibfnamefont
  {T.}~\bibnamefont {Kopp}}, \bibinfo {author} {\bibfnamefont {A.-S.}\
  \bibnamefont {R{\"{u}}etschi}}, \bibinfo {author} {\bibfnamefont
  {D.}~\bibnamefont {Jaccard}}, \bibinfo {author} {\bibfnamefont
  {M.}~\bibnamefont {Gabay}}, \bibinfo {author} {\bibfnamefont {D.~A.}\
  \bibnamefont {Muller}}, \bibinfo {author} {\bibfnamefont {J.-M.}\
  \bibnamefont {Triscone}}, \ and\ \bibinfo {author} {\bibfnamefont
  {J.}~\bibnamefont {Mannhart}},\ }\href {\doibase 10.1126/science.1146006}
  {\bibfield  {journal} {\bibinfo  {journal} {Science}\ }\textbf {\bibinfo
  {volume} {317}},\ \bibinfo {pages} {1196} (\bibinfo {year}
  {2007})}\BibitemShut {NoStop}%
\bibitem [{\citenamefont {Caviglia}\ \emph {et~al.}(2008)\citenamefont
  {Caviglia}, \citenamefont {Gariglio}, \citenamefont {Reyren}, \citenamefont
  {Jaccard}, \citenamefont {Schneider}, \citenamefont {Gabay}, \citenamefont
  {Thiel}, \citenamefont {Hammerl}, \citenamefont {Mannhart},\ and\
  \citenamefont {Triscone}}]{Caviglia2008}%
  \BibitemOpen
  \bibfield  {author} {\bibinfo {author} {\bibfnamefont {A.~D.}\ \bibnamefont
  {Caviglia}}, \bibinfo {author} {\bibfnamefont {S.}~\bibnamefont {Gariglio}},
  \bibinfo {author} {\bibfnamefont {N.}~\bibnamefont {Reyren}}, \bibinfo
  {author} {\bibfnamefont {D.}~\bibnamefont {Jaccard}}, \bibinfo {author}
  {\bibfnamefont {T.}~\bibnamefont {Schneider}}, \bibinfo {author}
  {\bibfnamefont {M.}~\bibnamefont {Gabay}}, \bibinfo {author} {\bibfnamefont
  {S.}~\bibnamefont {Thiel}}, \bibinfo {author} {\bibfnamefont
  {G.}~\bibnamefont {Hammerl}}, \bibinfo {author} {\bibfnamefont
  {J.}~\bibnamefont {Mannhart}}, \ and\ \bibinfo {author} {\bibfnamefont
  {J.-M.}\ \bibnamefont {Triscone}},\ }\href {\doibase 10.1038/nature07576}
  {\bibfield  {journal} {\bibinfo  {journal} {Nature}\ }\textbf {\bibinfo
  {volume} {456}},\ \bibinfo {pages} {624} (\bibinfo {year}
  {2008})}\BibitemShut {NoStop}%
\bibitem [{\citenamefont {Li}\ \emph {et~al.}(2011)\citenamefont {Li},
  \citenamefont {Richter}, \citenamefont {Mannhart},\ and\ \citenamefont
  {Ashoori}}]{Li2011}%
  \BibitemOpen
  \bibfield  {author} {\bibinfo {author} {\bibfnamefont {L.}~\bibnamefont
  {Li}}, \bibinfo {author} {\bibfnamefont {C.}~\bibnamefont {Richter}},
  \bibinfo {author} {\bibfnamefont {J.}~\bibnamefont {Mannhart}}, \ and\
  \bibinfo {author} {\bibfnamefont {R.~C.}\ \bibnamefont {Ashoori}},\ }\href
  {\doibase 10.1038/nphys2080} {\bibfield  {journal} {\bibinfo  {journal}
  {Nature Physics}\ }\textbf {\bibinfo {volume} {7}},\ \bibinfo {pages} {762}
  (\bibinfo {year} {2011})}\BibitemShut {NoStop}%
\bibitem [{\citenamefont {Caviglia}\ \emph {et~al.}(2010)\citenamefont
  {Caviglia}, \citenamefont {Gabay}, \citenamefont {Gariglio}, \citenamefont
  {Reyren}, \citenamefont {Cancellieri},\ and\ \citenamefont
  {Triscone}}]{Caviglia2010}%
  \BibitemOpen
  \bibfield  {author} {\bibinfo {author} {\bibfnamefont {A.~D.}\ \bibnamefont
  {Caviglia}}, \bibinfo {author} {\bibfnamefont {M.}~\bibnamefont {Gabay}},
  \bibinfo {author} {\bibfnamefont {S.}~\bibnamefont {Gariglio}}, \bibinfo
  {author} {\bibfnamefont {N.}~\bibnamefont {Reyren}}, \bibinfo {author}
  {\bibfnamefont {C.}~\bibnamefont {Cancellieri}}, \ and\ \bibinfo {author}
  {\bibfnamefont {J.-M.}\ \bibnamefont {Triscone}},\ }\href {\doibase
  10.1103/PhysRevLett.104.126803} {\bibfield  {journal} {\bibinfo  {journal}
  {Physical Review Letters}\ }\textbf {\bibinfo {volume} {104}},\ \bibinfo
  {pages} {126803} (\bibinfo {year} {2010})}\BibitemShut {NoStop}%
\bibitem [{\citenamefont {F{\^{e}}te}\ \emph {et~al.}(2014)\citenamefont
  {F{\^{e}}te}, \citenamefont {Gariglio}, \citenamefont {Berthod},
  \citenamefont {Li}, \citenamefont {Stornaiuolo}, \citenamefont {Gabay},\ and\
  \citenamefont {Triscone}}]{Fete2014}%
  \BibitemOpen
  \bibfield  {author} {\bibinfo {author} {\bibfnamefont {A.}~\bibnamefont
  {F{\^{e}}te}}, \bibinfo {author} {\bibfnamefont {S.}~\bibnamefont
  {Gariglio}}, \bibinfo {author} {\bibfnamefont {C.}~\bibnamefont {Berthod}},
  \bibinfo {author} {\bibfnamefont {D.}~\bibnamefont {Li}}, \bibinfo {author}
  {\bibfnamefont {D.}~\bibnamefont {Stornaiuolo}}, \bibinfo {author}
  {\bibfnamefont {M.}~\bibnamefont {Gabay}}, \ and\ \bibinfo {author}
  {\bibfnamefont {J.-M.}\ \bibnamefont {Triscone}},\ }\href {\doibase
  10.1088/1367-2630/16/11/112002} {\bibfield  {journal} {\bibinfo  {journal}
  {New Journal of Physics}\ }\textbf {\bibinfo {volume} {16}},\ \bibinfo
  {pages} {112002} (\bibinfo {year} {2014})},\ \Eprint
  {http://arxiv.org/abs/1407.4925} {1407.4925} \BibitemShut {NoStop}%
\bibitem [{\citenamefont {Hurand}\ \emph {et~al.}(2015)\citenamefont {Hurand},
  \citenamefont {Jouan}, \citenamefont {Feuillet-Palma}, \citenamefont {Singh},
  \citenamefont {Biscaras}, \citenamefont {Lesne}, \citenamefont {Reyren},
  \citenamefont {Barth{\'{e}}l{\'{e}}my}, \citenamefont {Bibes}, \citenamefont
  {Villegas}, \citenamefont {Ulysse}, \citenamefont {Lafosse}, \citenamefont
  {Pannetier-Lecoeur}, \citenamefont {Caprara}, \citenamefont {Grilli},
  \citenamefont {Lesueur},\ and\ \citenamefont {Bergeal}}]{Hurand2015}%
  \BibitemOpen
  \bibfield  {author} {\bibinfo {author} {\bibfnamefont {S.}~\bibnamefont
  {Hurand}}, \bibinfo {author} {\bibfnamefont {A.}~\bibnamefont {Jouan}},
  \bibinfo {author} {\bibfnamefont {C.}~\bibnamefont {Feuillet-Palma}},
  \bibinfo {author} {\bibfnamefont {G.}~\bibnamefont {Singh}}, \bibinfo
  {author} {\bibfnamefont {J.}~\bibnamefont {Biscaras}}, \bibinfo {author}
  {\bibfnamefont {E.}~\bibnamefont {Lesne}}, \bibinfo {author} {\bibfnamefont
  {N.}~\bibnamefont {Reyren}}, \bibinfo {author} {\bibfnamefont
  {A.}~\bibnamefont {Barth{\'{e}}l{\'{e}}my}}, \bibinfo {author} {\bibfnamefont
  {M.}~\bibnamefont {Bibes}}, \bibinfo {author} {\bibfnamefont {J.~E.}\
  \bibnamefont {Villegas}}, \bibinfo {author} {\bibfnamefont {C.}~\bibnamefont
  {Ulysse}}, \bibinfo {author} {\bibfnamefont {X.}~\bibnamefont {Lafosse}},
  \bibinfo {author} {\bibfnamefont {M.}~\bibnamefont {Pannetier-Lecoeur}},
  \bibinfo {author} {\bibfnamefont {S.}~\bibnamefont {Caprara}}, \bibinfo
  {author} {\bibfnamefont {M.}~\bibnamefont {Grilli}}, \bibinfo {author}
  {\bibfnamefont {J.}~\bibnamefont {Lesueur}}, \ and\ \bibinfo {author}
  {\bibfnamefont {N.}~\bibnamefont {Bergeal}},\ }\href {\doibase
  10.1038/srep12751} {\bibfield  {journal} {\bibinfo  {journal} {Scientific
  Reports}\ }\textbf {\bibinfo {volume} {5}},\ \bibinfo {pages} {12751}
  (\bibinfo {year} {2015})}\BibitemShut {NoStop}%
\bibitem [{\citenamefont {Liu}\ \emph {et~al.}(2013)\citenamefont {Liu},
  \citenamefont {Li}, \citenamefont {L{\"{u}}}, \citenamefont {Huang},
  \citenamefont {Huang}, \citenamefont {Zeng}, \citenamefont {Qiu},
  \citenamefont {Huang}, \citenamefont {Annadi}, \citenamefont {Chen},
  \citenamefont {Coey}, \citenamefont {Venkatesan},\ and\ \citenamefont
  {Ariando}}]{Liu2013}%
  \BibitemOpen
  \bibfield  {author} {\bibinfo {author} {\bibfnamefont {Z.~Q.}\ \bibnamefont
  {Liu}}, \bibinfo {author} {\bibfnamefont {C.~J.}\ \bibnamefont {Li}},
  \bibinfo {author} {\bibfnamefont {W.~M.}\ \bibnamefont {L{\"{u}}}}, \bibinfo
  {author} {\bibfnamefont {X.~H.}\ \bibnamefont {Huang}}, \bibinfo {author}
  {\bibfnamefont {Z.}~\bibnamefont {Huang}}, \bibinfo {author} {\bibfnamefont
  {S.~W.}\ \bibnamefont {Zeng}}, \bibinfo {author} {\bibfnamefont {X.~P.}\
  \bibnamefont {Qiu}}, \bibinfo {author} {\bibfnamefont {L.~S.}\ \bibnamefont
  {Huang}}, \bibinfo {author} {\bibfnamefont {A.}~\bibnamefont {Annadi}},
  \bibinfo {author} {\bibfnamefont {J.~S.}\ \bibnamefont {Chen}}, \bibinfo
  {author} {\bibfnamefont {J.~M.~D.}\ \bibnamefont {Coey}}, \bibinfo {author}
  {\bibfnamefont {T.}~\bibnamefont {Venkatesan}}, \ and\ \bibinfo {author}
  {\bibnamefont {Ariando}},\ }\href {\doibase 10.1103/PhysRevX.3.021010}
  {\bibfield  {journal} {\bibinfo  {journal} {Physical Review X}\ }\textbf
  {\bibinfo {volume} {3}},\ \bibinfo {pages} {021010} (\bibinfo {year}
  {2013})}\BibitemShut {NoStop}%
\bibitem [{\citenamefont {Ueno}\ \emph {et~al.}(2008)\citenamefont {Ueno},
  \citenamefont {Nakamura}, \citenamefont {Shimotani}, \citenamefont {Ohtomo},
  \citenamefont {Kimura}, \citenamefont {Nojima}, \citenamefont {Aoki},
  \citenamefont {Iwasa},\ and\ \citenamefont {Kawasaki}}]{Ueno2008}%
  \BibitemOpen
  \bibfield  {author} {\bibinfo {author} {\bibfnamefont {K.}~\bibnamefont
  {Ueno}}, \bibinfo {author} {\bibfnamefont {S.}~\bibnamefont {Nakamura}},
  \bibinfo {author} {\bibfnamefont {H.}~\bibnamefont {Shimotani}}, \bibinfo
  {author} {\bibfnamefont {A.}~\bibnamefont {Ohtomo}}, \bibinfo {author}
  {\bibfnamefont {N.}~\bibnamefont {Kimura}}, \bibinfo {author} {\bibfnamefont
  {T.}~\bibnamefont {Nojima}}, \bibinfo {author} {\bibfnamefont
  {H.}~\bibnamefont {Aoki}}, \bibinfo {author} {\bibfnamefont {Y.}~\bibnamefont
  {Iwasa}}, \ and\ \bibinfo {author} {\bibfnamefont {M.}~\bibnamefont
  {Kawasaki}},\ }\href {\doibase 10.1038/nmat2298} {\bibfield  {journal}
  {\bibinfo  {journal} {Nature materials}\ }\textbf {\bibinfo {volume} {7}},\
  \bibinfo {pages} {855} (\bibinfo {year} {2008})}\BibitemShut {NoStop}%
\bibitem [{\citenamefont {Meevasana}\ \emph {et~al.}(2011)\citenamefont
  {Meevasana}, \citenamefont {King}, \citenamefont {He}, \citenamefont {Mo},
  \citenamefont {Hashimoto}, \citenamefont {Tamai}, \citenamefont
  {Songsiriritthigul}, \citenamefont {Baumberger},\ and\ \citenamefont
  {Shen}}]{Meevasana2011a}%
  \BibitemOpen
  \bibfield  {author} {\bibinfo {author} {\bibfnamefont {W.}~\bibnamefont
  {Meevasana}}, \bibinfo {author} {\bibfnamefont {P.~D.~C.}\ \bibnamefont
  {King}}, \bibinfo {author} {\bibfnamefont {R.~H.}\ \bibnamefont {He}},
  \bibinfo {author} {\bibfnamefont {S.-K.}\ \bibnamefont {Mo}}, \bibinfo
  {author} {\bibfnamefont {M.}~\bibnamefont {Hashimoto}}, \bibinfo {author}
  {\bibfnamefont {A.}~\bibnamefont {Tamai}}, \bibinfo {author} {\bibfnamefont
  {P.}~\bibnamefont {Songsiriritthigul}}, \bibinfo {author} {\bibfnamefont
  {F.}~\bibnamefont {Baumberger}}, \ and\ \bibinfo {author} {\bibfnamefont
  {Z.-X.}\ \bibnamefont {Shen}},\ }\href {\doibase 10.1038/nmat2943} {\bibfield
   {journal} {\bibinfo  {journal} {Nature Materials}\ }\textbf {\bibinfo
  {volume} {10}},\ \bibinfo {pages} {114} (\bibinfo {year} {2011})}\BibitemShut
  {NoStop}%
\bibitem [{\citenamefont {McKeown~Walker}\ \emph {et~al.}(2015)\citenamefont
  {McKeown~Walker}, \citenamefont {Bruno}, \citenamefont {Wang}, \citenamefont
  {de~la Torre}, \citenamefont {Ricc{\'{o}}}, \citenamefont {Tamai},
  \citenamefont {Kim}, \citenamefont {Hoesch}, \citenamefont {Shi},
  \citenamefont {Bahramy}, \citenamefont {King},\ and\ \citenamefont
  {Baumberger}}]{McKeownWalker2015}%
  \BibitemOpen
  \bibfield  {author} {\bibinfo {author} {\bibfnamefont {S.}~\bibnamefont
  {McKeown~Walker}}, \bibinfo {author} {\bibfnamefont {F.~Y.}\ \bibnamefont
  {Bruno}}, \bibinfo {author} {\bibfnamefont {Z.}~\bibnamefont {Wang}},
  \bibinfo {author} {\bibfnamefont {A.}~\bibnamefont {de~la Torre}}, \bibinfo
  {author} {\bibfnamefont {S.}~\bibnamefont {Ricc{\'{o}}}}, \bibinfo {author}
  {\bibfnamefont {A.}~\bibnamefont {Tamai}}, \bibinfo {author} {\bibfnamefont
  {T.~K.}\ \bibnamefont {Kim}}, \bibinfo {author} {\bibfnamefont
  {M.}~\bibnamefont {Hoesch}}, \bibinfo {author} {\bibfnamefont
  {M.}~\bibnamefont {Shi}}, \bibinfo {author} {\bibfnamefont {M.~S.}\
  \bibnamefont {Bahramy}}, \bibinfo {author} {\bibfnamefont {P.~D.~C.}\
  \bibnamefont {King}}, \ and\ \bibinfo {author} {\bibfnamefont
  {F.}~\bibnamefont {Baumberger}},\ }\href {\doibase 10.1002/adma.201501556}
  {\bibfield  {journal} {\bibinfo  {journal} {Advanced Materials}\ }\textbf
  {\bibinfo {volume} {27}},\ \bibinfo {pages} {3894} (\bibinfo {year}
  {2015})}\BibitemShut {NoStop}%
\bibitem [{\citenamefont {R{\"{o}}del}\ \emph {et~al.}(2016)\citenamefont
  {R{\"{o}}del}, \citenamefont {Fortuna}, \citenamefont {Sengupta},
  \citenamefont {Frantzeskakis}, \citenamefont {F{\`{e}}vre}, \citenamefont
  {Bertran}, \citenamefont {Mercey}, \citenamefont {Matzen}, \citenamefont
  {Agnus}, \citenamefont {Maroutian}, \citenamefont {Lecoeur},\ and\
  \citenamefont {Santander-Syro}}]{Rodel2016}%
  \BibitemOpen
  \bibfield  {author} {\bibinfo {author} {\bibfnamefont {T.~C.}\ \bibnamefont
  {R{\"{o}}del}}, \bibinfo {author} {\bibfnamefont {F.}~\bibnamefont
  {Fortuna}}, \bibinfo {author} {\bibfnamefont {S.}~\bibnamefont {Sengupta}},
  \bibinfo {author} {\bibfnamefont {E.}~\bibnamefont {Frantzeskakis}}, \bibinfo
  {author} {\bibfnamefont {P.~L.}\ \bibnamefont {F{\`{e}}vre}}, \bibinfo
  {author} {\bibfnamefont {F.}~\bibnamefont {Bertran}}, \bibinfo {author}
  {\bibfnamefont {B.}~\bibnamefont {Mercey}}, \bibinfo {author} {\bibfnamefont
  {S.}~\bibnamefont {Matzen}}, \bibinfo {author} {\bibfnamefont
  {G.}~\bibnamefont {Agnus}}, \bibinfo {author} {\bibfnamefont
  {T.}~\bibnamefont {Maroutian}}, \bibinfo {author} {\bibfnamefont
  {P.}~\bibnamefont {Lecoeur}}, \ and\ \bibinfo {author} {\bibfnamefont
  {A.~F.}\ \bibnamefont {Santander-Syro}},\ }\href {\doibase
  10.1002/adma.201505021} {\bibfield  {journal} {\bibinfo  {journal} {Advanced
  Materials}\ } (\bibinfo {year} {2016}),\ 10.1002/adma.201505021}\BibitemShut
  {NoStop}%
\bibitem [{\citenamefont {Stengel}(2011)}]{Stengel2011}%
  \BibitemOpen
  \bibfield  {author} {\bibinfo {author} {\bibfnamefont {M.}~\bibnamefont
  {Stengel}},\ }\href {http://link.aps.org/doi/10.1103/PhysRevLett.106.136803}
  {\bibfield  {journal} {\bibinfo  {journal} {Phys. Rev. Lett.}\ }\textbf
  {\bibinfo {volume} {106}},\ \bibinfo {pages} {136803} (\bibinfo {year}
  {2011})}\BibitemShut {NoStop}%
\bibitem [{\citenamefont {Santander-Syro}\ \emph {et~al.}(2011)\citenamefont
  {Santander-Syro}, \citenamefont {Copie}, \citenamefont {Kondo}, \citenamefont
  {Fortuna}, \citenamefont {Pailh{\`{e}}s}, \citenamefont {Weht}, \citenamefont
  {Qiu}, \citenamefont {Bertran}, \citenamefont {Nicolaou}, \citenamefont
  {Taleb-Ibrahimi}, \citenamefont {{Le F{\`{e}}vre}}, \citenamefont {Herranz},
  \citenamefont {Bibes}, \citenamefont {Reyren}, \citenamefont {Apertet},
  \citenamefont {Lecoeur}, \citenamefont {Barth{\'{e}}l{\'{e}}my},\ and\
  \citenamefont {Rozenberg}}]{Santander-Syro2011a}%
  \BibitemOpen
  \bibfield  {author} {\bibinfo {author} {\bibfnamefont {A.~F.}\ \bibnamefont
  {Santander-Syro}}, \bibinfo {author} {\bibfnamefont {O.}~\bibnamefont
  {Copie}}, \bibinfo {author} {\bibfnamefont {T.}~\bibnamefont {Kondo}},
  \bibinfo {author} {\bibfnamefont {F.}~\bibnamefont {Fortuna}}, \bibinfo
  {author} {\bibfnamefont {S.}~\bibnamefont {Pailh{\`{e}}s}}, \bibinfo {author}
  {\bibfnamefont {R.}~\bibnamefont {Weht}}, \bibinfo {author} {\bibfnamefont
  {X.~G.}\ \bibnamefont {Qiu}}, \bibinfo {author} {\bibfnamefont
  {F.}~\bibnamefont {Bertran}}, \bibinfo {author} {\bibfnamefont
  {A.}~\bibnamefont {Nicolaou}}, \bibinfo {author} {\bibfnamefont
  {A.}~\bibnamefont {Taleb-Ibrahimi}}, \bibinfo {author} {\bibfnamefont
  {P.}~\bibnamefont {{Le F{\`{e}}vre}}}, \bibinfo {author} {\bibfnamefont
  {G.}~\bibnamefont {Herranz}}, \bibinfo {author} {\bibfnamefont
  {M.}~\bibnamefont {Bibes}}, \bibinfo {author} {\bibfnamefont
  {N.}~\bibnamefont {Reyren}}, \bibinfo {author} {\bibfnamefont
  {Y.}~\bibnamefont {Apertet}}, \bibinfo {author} {\bibfnamefont
  {P.}~\bibnamefont {Lecoeur}}, \bibinfo {author} {\bibfnamefont
  {A.}~\bibnamefont {Barth{\'{e}}l{\'{e}}my}}, \ and\ \bibinfo {author}
  {\bibfnamefont {M.~J.}\ \bibnamefont {Rozenberg}},\ }\href {\doibase
  10.1038/nature09720} {\bibfield  {journal} {\bibinfo  {journal} {Nature}\
  }\textbf {\bibinfo {volume} {469}},\ \bibinfo {pages} {189} (\bibinfo {year}
  {2011})}\BibitemShut {NoStop}%
\bibitem [{\citenamefont {Joshua}\ \emph {et~al.}(2012)\citenamefont {Joshua},
  \citenamefont {Pecker}, \citenamefont {Ruhman}, \citenamefont {Altman},\ and\
  \citenamefont {Ilani}}]{Joshua2012}%
  \BibitemOpen
  \bibfield  {author} {\bibinfo {author} {\bibfnamefont {A.}~\bibnamefont
  {Joshua}}, \bibinfo {author} {\bibfnamefont {S.}~\bibnamefont {Pecker}},
  \bibinfo {author} {\bibfnamefont {J.}~\bibnamefont {Ruhman}}, \bibinfo
  {author} {\bibfnamefont {E.}~\bibnamefont {Altman}}, \ and\ \bibinfo {author}
  {\bibfnamefont {S.}~\bibnamefont {Ilani}},\ }\href {\doibase
  10.1038/ncomms2116} {\bibfield  {journal} {\bibinfo  {journal} {Nature
  Communications}\ }\textbf {\bibinfo {volume} {3}},\ \bibinfo {pages} {1129}
  (\bibinfo {year} {2012})}\BibitemShut {NoStop}%
\bibitem [{\citenamefont {King}\ \emph {et~al.}(2014)\citenamefont {King},
  \citenamefont {{McKeown Walker}}, \citenamefont {Tamai}, \citenamefont {de~la
  Torre}, \citenamefont {Eknapakul}, \citenamefont {Buaphet}, \citenamefont
  {Mo}, \citenamefont {Meevasana}, \citenamefont {Bahramy},\ and\ \citenamefont
  {Baumberger}}]{King2014}%
  \BibitemOpen
  \bibfield  {author} {\bibinfo {author} {\bibfnamefont {P.~D.~C.}\
  \bibnamefont {King}}, \bibinfo {author} {\bibfnamefont {S.}~\bibnamefont
  {{McKeown Walker}}}, \bibinfo {author} {\bibfnamefont {A.}~\bibnamefont
  {Tamai}}, \bibinfo {author} {\bibfnamefont {A.}~\bibnamefont {de~la Torre}},
  \bibinfo {author} {\bibfnamefont {T.}~\bibnamefont {Eknapakul}}, \bibinfo
  {author} {\bibfnamefont {P.}~\bibnamefont {Buaphet}}, \bibinfo {author}
  {\bibfnamefont {S.-K.}\ \bibnamefont {Mo}}, \bibinfo {author} {\bibfnamefont
  {W.}~\bibnamefont {Meevasana}}, \bibinfo {author} {\bibfnamefont {M.~S.}\
  \bibnamefont {Bahramy}}, \ and\ \bibinfo {author} {\bibfnamefont
  {F.}~\bibnamefont {Baumberger}},\ }\href {\doibase 10.1038/ncomms4414}
  {\bibfield  {journal} {\bibinfo  {journal} {Nature Communications}\ }\textbf
  {\bibinfo {volume} {5}},\ \bibinfo {pages} {3414} (\bibinfo {year}
  {2014})}\BibitemShut {NoStop}%
\bibitem [{\citenamefont {Salluzzo}\ \emph {et~al.}(2009)\citenamefont
  {Salluzzo}, \citenamefont {Cezar}, \citenamefont {Brookes}, \citenamefont
  {Bisogni}, \citenamefont {{De Luca}}, \citenamefont {Richter}, \citenamefont
  {Thiel}, \citenamefont {Mannhart}, \citenamefont {Huijben}, \citenamefont
  {Brinkman}, \citenamefont {Rijnders},\ and\ \citenamefont
  {Ghiringhelli}}]{Salluzzo2009}%
  \BibitemOpen
  \bibfield  {author} {\bibinfo {author} {\bibfnamefont {M.}~\bibnamefont
  {Salluzzo}}, \bibinfo {author} {\bibfnamefont {J.~C.}\ \bibnamefont {Cezar}},
  \bibinfo {author} {\bibfnamefont {N.~B.}\ \bibnamefont {Brookes}}, \bibinfo
  {author} {\bibfnamefont {V.}~\bibnamefont {Bisogni}}, \bibinfo {author}
  {\bibfnamefont {G.~M.}\ \bibnamefont {{De Luca}}}, \bibinfo {author}
  {\bibfnamefont {C.}~\bibnamefont {Richter}}, \bibinfo {author} {\bibfnamefont
  {S.}~\bibnamefont {Thiel}}, \bibinfo {author} {\bibfnamefont
  {J.}~\bibnamefont {Mannhart}}, \bibinfo {author} {\bibfnamefont
  {M.}~\bibnamefont {Huijben}}, \bibinfo {author} {\bibfnamefont
  {A.}~\bibnamefont {Brinkman}}, \bibinfo {author} {\bibfnamefont
  {G.}~\bibnamefont {Rijnders}}, \ and\ \bibinfo {author} {\bibfnamefont
  {G.}~\bibnamefont {Ghiringhelli}},\ }\href {\doibase
  10.1103/PhysRevLett.102.166804} {\bibfield  {journal} {\bibinfo  {journal}
  {Physical Review Letters}\ }\textbf {\bibinfo {volume} {102}},\ \bibinfo
  {pages} {166804} (\bibinfo {year} {2009})}\BibitemShut {NoStop}%
\bibitem [{\citenamefont {Cancellieri}\ \emph {et~al.}(2014)\citenamefont
  {Cancellieri}, \citenamefont {Reinle-Schmitt}, \citenamefont {Kobayashi},
  \citenamefont {Strocov}, \citenamefont {Willmott}, \citenamefont {Fontaine},
  \citenamefont {Ghosez}, \citenamefont {Filippetti}, \citenamefont {Delugas},\
  and\ \citenamefont {Fiorentini}}]{Cancellieri2014}%
  \BibitemOpen
  \bibfield  {author} {\bibinfo {author} {\bibfnamefont {C.}~\bibnamefont
  {Cancellieri}}, \bibinfo {author} {\bibfnamefont {M.~L.}\ \bibnamefont
  {Reinle-Schmitt}}, \bibinfo {author} {\bibfnamefont {M.}~\bibnamefont
  {Kobayashi}}, \bibinfo {author} {\bibfnamefont {V.~N.}\ \bibnamefont
  {Strocov}}, \bibinfo {author} {\bibfnamefont {P.~R.}\ \bibnamefont
  {Willmott}}, \bibinfo {author} {\bibfnamefont {D.}~\bibnamefont {Fontaine}},
  \bibinfo {author} {\bibfnamefont {P.}~\bibnamefont {Ghosez}}, \bibinfo
  {author} {\bibfnamefont {A.}~\bibnamefont {Filippetti}}, \bibinfo {author}
  {\bibfnamefont {P.}~\bibnamefont {Delugas}}, \ and\ \bibinfo {author}
  {\bibfnamefont {V.}~\bibnamefont {Fiorentini}},\ }\href {\doibase
  10.1103/PhysRevB.89.121412} {\bibfield  {journal} {\bibinfo  {journal}
  {Physical Review B}\ }\textbf {\bibinfo {volume} {89}},\ \bibinfo {pages}
  {121412} (\bibinfo {year} {2014})}\BibitemShut {NoStop}%
\bibitem [{\citenamefont {Zhong}\ \emph {et~al.}(2013)\citenamefont {Zhong},
  \citenamefont {T{\'{o}}th},\ and\ \citenamefont {Held}}]{Zhong2013b}%
  \BibitemOpen
  \bibfield  {author} {\bibinfo {author} {\bibfnamefont {Z.}~\bibnamefont
  {Zhong}}, \bibinfo {author} {\bibfnamefont {A.}~\bibnamefont {T{\'{o}}th}}, \
  and\ \bibinfo {author} {\bibfnamefont {K.}~\bibnamefont {Held}},\ }\href
  {\doibase 10.1103/PhysRevB.87.161102} {\bibfield  {journal} {\bibinfo
  {journal} {Physical Review B}\ }\textbf {\bibinfo {volume} {87}},\ \bibinfo
  {pages} {161102} (\bibinfo {year} {2013})}\BibitemShut {NoStop}%
\bibitem [{\citenamefont {Khalsa}\ \emph {et~al.}(2013)\citenamefont {Khalsa},
  \citenamefont {Lee},\ and\ \citenamefont {MacDonald}}]{Khalsa2013a}%
  \BibitemOpen
  \bibfield  {author} {\bibinfo {author} {\bibfnamefont {G.}~\bibnamefont
  {Khalsa}}, \bibinfo {author} {\bibfnamefont {B.}~\bibnamefont {Lee}}, \ and\
  \bibinfo {author} {\bibfnamefont {A.~H.}\ \bibnamefont {MacDonald}},\ }\href
  {\doibase 10.1103/PhysRevB.88.041302} {\bibfield  {journal} {\bibinfo
  {journal} {Physical Review B}\ }\textbf {\bibinfo {volume} {88}},\ \bibinfo
  {pages} {041302} (\bibinfo {year} {2013})}\BibitemShut {NoStop}%
\bibitem [{\citenamefont {Kim}\ \emph {et~al.}(2014)\citenamefont {Kim},
  \citenamefont {Kang}, \citenamefont {Go},\ and\ \citenamefont
  {Han}}]{Kim2014a}%
  \BibitemOpen
  \bibfield  {author} {\bibinfo {author} {\bibfnamefont {P.}~\bibnamefont
  {Kim}}, \bibinfo {author} {\bibfnamefont {K.~T.}\ \bibnamefont {Kang}},
  \bibinfo {author} {\bibfnamefont {G.}~\bibnamefont {Go}}, \ and\ \bibinfo
  {author} {\bibfnamefont {J.~H.}\ \bibnamefont {Han}},\ }\href {\doibase
  10.1103/PhysRevB.90.205423} {\bibfield  {journal} {\bibinfo  {journal}
  {Physical Review B}\ }\textbf {\bibinfo {volume} {90}},\ \bibinfo {pages}
  {205423} (\bibinfo {year} {2014})}\BibitemShut {NoStop}%
\bibitem [{\citenamefont {Kim}\ \emph {et~al.}(2013)\citenamefont {Kim},
  \citenamefont {Lutchyn},\ and\ \citenamefont {Nayak}}]{Kim2013a}%
  \BibitemOpen
  \bibfield  {author} {\bibinfo {author} {\bibfnamefont {Y.}~\bibnamefont
  {Kim}}, \bibinfo {author} {\bibfnamefont {R.~M.}\ \bibnamefont {Lutchyn}}, \
  and\ \bibinfo {author} {\bibfnamefont {C.}~\bibnamefont {Nayak}},\ }\href
  {\doibase 10.1103/PhysRevB.87.245121} {\bibfield  {journal} {\bibinfo
  {journal} {Physical Review B - Condensed Matter and Materials Physics}\
  }\textbf {\bibinfo {volume} {87}},\ \bibinfo {pages} {245121} (\bibinfo
  {year} {2013})}\BibitemShut {NoStop}%
\bibitem [{\citenamefont {Altmeyer}\ \emph {et~al.}(2015)\citenamefont
  {Altmeyer}, \citenamefont {Jeschke}, \citenamefont {Hijano-Cubelos},
  \citenamefont {Martins}, \citenamefont {Lechermann}, \citenamefont
  {Koepernik}, \citenamefont {Santander-Syro}, \citenamefont {Rozenberg},
  \citenamefont {Valenti},\ and\ \citenamefont {Gabay}}]{Altmeyer2015}%
  \BibitemOpen
  \bibfield  {author} {\bibinfo {author} {\bibfnamefont {M.}~\bibnamefont
  {Altmeyer}}, \bibinfo {author} {\bibfnamefont {H.~O.}\ \bibnamefont
  {Jeschke}}, \bibinfo {author} {\bibfnamefont {O.}~\bibnamefont
  {Hijano-Cubelos}}, \bibinfo {author} {\bibfnamefont {C.}~\bibnamefont
  {Martins}}, \bibinfo {author} {\bibfnamefont {F.}~\bibnamefont {Lechermann}},
  \bibinfo {author} {\bibfnamefont {K.}~\bibnamefont {Koepernik}}, \bibinfo
  {author} {\bibfnamefont {A.}~\bibnamefont {Santander-Syro}}, \bibinfo
  {author} {\bibfnamefont {M.~J.}\ \bibnamefont {Rozenberg}}, \bibinfo {author}
  {\bibfnamefont {R.}~\bibnamefont {Valenti}}, \ and\ \bibinfo {author}
  {\bibfnamefont {M.}~\bibnamefont {Gabay}},\ }\href
  {http://arxiv.org/abs/1511.08614} {\  (\bibinfo {year} {2015})},\ \Eprint
  {http://arxiv.org/abs/1511.08614} {arXiv:1511.08614} \BibitemShut {NoStop}%
\bibitem [{\citenamefont {Garcia-Castro}\ \emph {et~al.}(2015)\citenamefont
  {Garcia-Castro}, \citenamefont {Vergniory}, \citenamefont {Bousquet},\ and\
  \citenamefont {Romero}}]{Garcia-Castro2015b}%
  \BibitemOpen
  \bibfield  {author} {\bibinfo {author} {\bibfnamefont {A.~C.}\ \bibnamefont
  {Garcia-Castro}}, \bibinfo {author} {\bibfnamefont {M.~G.}\ \bibnamefont
  {Vergniory}}, \bibinfo {author} {\bibfnamefont {E.}~\bibnamefont {Bousquet}},
  \ and\ \bibinfo {author} {\bibfnamefont {A.~H.}\ \bibnamefont {Romero}},\
  }\href {http://arxiv.org/abs/1511.08079} {\  (\bibinfo {year} {2015})},\
  \Eprint {http://arxiv.org/abs/1511.08079} {arXiv:1511.08079} \BibitemShut
  {NoStop}%
\bibitem [{\citenamefont {Nakamura}\ \emph {et~al.}(2012)\citenamefont
  {Nakamura}, \citenamefont {Koga},\ and\ \citenamefont
  {Kimura}}]{Nakamura2012}%
  \BibitemOpen
  \bibfield  {author} {\bibinfo {author} {\bibfnamefont {H.}~\bibnamefont
  {Nakamura}}, \bibinfo {author} {\bibfnamefont {T.}~\bibnamefont {Koga}}, \
  and\ \bibinfo {author} {\bibfnamefont {T.}~\bibnamefont {Kimura}},\ }\href
  {\doibase 10.1103/PhysRevLett.108.206601} {\bibfield  {journal} {\bibinfo
  {journal} {Physical Review Letters}\ }\textbf {\bibinfo {volume} {108}},\
  \bibinfo {pages} {206601} (\bibinfo {year} {2012})}\BibitemShut {NoStop}%
\bibitem [{\citenamefont {Liang}\ \emph {et~al.}(2015)\citenamefont {Liang},
  \citenamefont {Cheng}, \citenamefont {Wei}, \citenamefont {Luo},
  \citenamefont {Yu}, \citenamefont {Zeng},\ and\ \citenamefont
  {Zhang}}]{Liang2015}%
  \BibitemOpen
  \bibfield  {author} {\bibinfo {author} {\bibfnamefont {H.}~\bibnamefont
  {Liang}}, \bibinfo {author} {\bibfnamefont {L.}~\bibnamefont {Cheng}},
  \bibinfo {author} {\bibfnamefont {L.}~\bibnamefont {Wei}}, \bibinfo {author}
  {\bibfnamefont {Z.}~\bibnamefont {Luo}}, \bibinfo {author} {\bibfnamefont
  {G.}~\bibnamefont {Yu}}, \bibinfo {author} {\bibfnamefont {C.}~\bibnamefont
  {Zeng}}, \ and\ \bibinfo {author} {\bibfnamefont {Z.}~\bibnamefont {Zhang}},\
  }\href {\doibase 10.1103/PhysRevB.92.075309} {\bibfield  {journal} {\bibinfo
  {journal} {Phys. Rev. B}\ }\textbf {\bibinfo {volume} {92}},\ \bibinfo
  {pages} {075309} (\bibinfo {year} {2015})}\BibitemShut {NoStop}%
\bibitem [{\citenamefont {Santander-Syro}\ \emph {et~al.}(2014)\citenamefont
  {Santander-Syro}, \citenamefont {Fortuna}, \citenamefont {Bareille},
  \citenamefont {R{\"{o}}del}, \citenamefont {Landolt}, \citenamefont {Plumb},
  \citenamefont {Dil},\ and\ \citenamefont
  {Radovi{\'{c}}}}]{Santander-Syro2014}%
  \BibitemOpen
  \bibfield  {author} {\bibinfo {author} {\bibfnamefont {A.~F.}\ \bibnamefont
  {Santander-Syro}}, \bibinfo {author} {\bibfnamefont {F.}~\bibnamefont
  {Fortuna}}, \bibinfo {author} {\bibfnamefont {C.}~\bibnamefont {Bareille}},
  \bibinfo {author} {\bibfnamefont {T.~C.}\ \bibnamefont {R{\"{o}}del}},
  \bibinfo {author} {\bibfnamefont {G.}~\bibnamefont {Landolt}}, \bibinfo
  {author} {\bibfnamefont {N.~C.}\ \bibnamefont {Plumb}}, \bibinfo {author}
  {\bibfnamefont {J.~H.}\ \bibnamefont {Dil}}, \ and\ \bibinfo {author}
  {\bibfnamefont {M.}~\bibnamefont {Radovi{\'{c}}}},\ }\href {\doibase
  10.1038/nmat4107} {\bibfield  {journal} {\bibinfo  {journal} {Nature
  Materials}\ }\textbf {\bibinfo {volume} {13}},\ \bibinfo {pages} {1085}
  (\bibinfo {year} {2014})}\BibitemShut {NoStop}%
\bibitem [{\citenamefont {Ghosh}\ and\ \citenamefont
  {Manousakis}(2015)}]{Ghosh2015}%
  \BibitemOpen
  \bibfield  {author} {\bibinfo {author} {\bibfnamefont {S.~S.}\ \bibnamefont
  {Ghosh}}\ and\ \bibinfo {author} {\bibfnamefont {E.}~\bibnamefont
  {Manousakis}},\ }\href {http://arxiv.org/abs/1511.07495} {\  (\bibinfo {year}
  {2015})},\ \Eprint {http://arxiv.org/abs/1511.07495} {arXiv:1511.07495}
  \BibitemShut {NoStop}%
\bibitem [{\citenamefont {Lee}\ \emph {et~al.}(2011)\citenamefont {Lee},
  \citenamefont {Williams}, \citenamefont {Zhang}, \citenamefont {Frisbie},\
  and\ \citenamefont {Goldhaber-Gordon}}]{Lee2011c}%
  \BibitemOpen
  \bibfield  {author} {\bibinfo {author} {\bibfnamefont {M.}~\bibnamefont
  {Lee}}, \bibinfo {author} {\bibfnamefont {J.~R.}\ \bibnamefont {Williams}},
  \bibinfo {author} {\bibfnamefont {S.}~\bibnamefont {Zhang}}, \bibinfo
  {author} {\bibfnamefont {C.~D.}\ \bibnamefont {Frisbie}}, \ and\ \bibinfo
  {author} {\bibfnamefont {D.}~\bibnamefont {Goldhaber-Gordon}},\ }\href
  {\doibase 10.1103/PhysRevLett.107.256601} {\bibfield  {journal} {\bibinfo
  {journal} {Physical Review Letters}\ }\textbf {\bibinfo {volume} {107}},\
  \bibinfo {pages} {256601} (\bibinfo {year} {2011})}\BibitemShut {NoStop}%
\bibitem [{\citenamefont {Tamai}\ \emph {et~al.}(2013)\citenamefont {Tamai},
  \citenamefont {Meevasana}, \citenamefont {King}, \citenamefont {Nicholson},
  \citenamefont {de~la Torre}, \citenamefont {Rozbicki},\ and\ \citenamefont
  {Baumberger}}]{Tamai2013b}%
  \BibitemOpen
  \bibfield  {author} {\bibinfo {author} {\bibfnamefont {A.}~\bibnamefont
  {Tamai}}, \bibinfo {author} {\bibfnamefont {W.}~\bibnamefont {Meevasana}},
  \bibinfo {author} {\bibfnamefont {P.~D.~C.}\ \bibnamefont {King}}, \bibinfo
  {author} {\bibfnamefont {C.~W.}\ \bibnamefont {Nicholson}}, \bibinfo {author}
  {\bibfnamefont {A.}~\bibnamefont {de~la Torre}}, \bibinfo {author}
  {\bibfnamefont {E.}~\bibnamefont {Rozbicki}}, \ and\ \bibinfo {author}
  {\bibfnamefont {F.}~\bibnamefont {Baumberger}},\ }\href {\doibase
  10.1103/PhysRevB.87.075113} {\bibfield  {journal} {\bibinfo  {journal}
  {Physical Review B}\ }\textbf {\bibinfo {volume} {87}},\ \bibinfo {pages}
  {075113} (\bibinfo {year} {2013})}\BibitemShut {NoStop}%
\bibitem [{\citenamefont {Marchenko}\ \emph {et~al.}(2013)\citenamefont
  {Marchenko}, \citenamefont {Varykhalov}, \citenamefont {Scholz},
  \citenamefont {S{\'{a}}nchez-Barriga}, \citenamefont {Rader}, \citenamefont
  {Rybkina}, \citenamefont {Shikin}, \citenamefont {Seyller},\ and\
  \citenamefont {Bihlmayer}}]{Marchenko2013}%
  \BibitemOpen
  \bibfield  {author} {\bibinfo {author} {\bibfnamefont {D.}~\bibnamefont
  {Marchenko}}, \bibinfo {author} {\bibfnamefont {A.}~\bibnamefont
  {Varykhalov}}, \bibinfo {author} {\bibfnamefont {M.~R.}\ \bibnamefont
  {Scholz}}, \bibinfo {author} {\bibfnamefont {J.}~\bibnamefont
  {S{\'{a}}nchez-Barriga}}, \bibinfo {author} {\bibfnamefont {O.}~\bibnamefont
  {Rader}}, \bibinfo {author} {\bibfnamefont {A.}~\bibnamefont {Rybkina}},
  \bibinfo {author} {\bibfnamefont {A.~M.}\ \bibnamefont {Shikin}}, \bibinfo
  {author} {\bibfnamefont {T.}~\bibnamefont {Seyller}}, \ and\ \bibinfo
  {author} {\bibfnamefont {G.}~\bibnamefont {Bihlmayer}},\ }\href {\doibase
  10.1103/PhysRevB.88.075422} {\bibfield  {journal} {\bibinfo  {journal}
  {Physical Review B}\ }\textbf {\bibinfo {volume} {88}},\ \bibinfo {pages}
  {075422} (\bibinfo {year} {2013})}\BibitemShut {NoStop}%
\bibitem [{\citenamefont {Burnett}\ \emph {et~al.}(1994)\citenamefont
  {Burnett}, \citenamefont {Monroe},\ and\ \citenamefont
  {Dunning}}]{Burnett-RSI-94}%
  \BibitemOpen
  \bibfield  {author} {\bibinfo {author} {\bibfnamefont {G.~C.}\ \bibnamefont
  {Burnett}}, \bibinfo {author} {\bibfnamefont {T.~J.}\ \bibnamefont {Monroe}},
  \ and\ \bibinfo {author} {\bibfnamefont {F.~N.}\ \bibnamefont {Dunning}},\
  }\href@noop {} {\bibfield  {journal} {\bibinfo  {journal} {Rev. Sci.
  Instrum.}\ }\textbf {\bibinfo {volume} {65}},\ \bibinfo {pages} {1893}
  (\bibinfo {year} {1994})}\BibitemShut {NoStop}%
\bibitem [{\citenamefont {Bahramy}\ \emph {et~al.}(2012)\citenamefont
  {Bahramy}, \citenamefont {King}, \citenamefont {de~la Torre}, \citenamefont
  {Chang}, \citenamefont {Shi}, \citenamefont {Patthey}, \citenamefont
  {Balakrishnan}, \citenamefont {Hofmann}, \citenamefont {Arita}, \citenamefont
  {Nagaosa},\ and\ \citenamefont {Baumberger}}]{Bahramy2012}%
  \BibitemOpen
  \bibfield  {author} {\bibinfo {author} {\bibfnamefont {M.~S.}\ \bibnamefont
  {Bahramy}}, \bibinfo {author} {\bibfnamefont {P.~D.~C.}\ \bibnamefont
  {King}}, \bibinfo {author} {\bibfnamefont {A.}~\bibnamefont {de~la Torre}},
  \bibinfo {author} {\bibfnamefont {J.}~\bibnamefont {Chang}}, \bibinfo
  {author} {\bibfnamefont {M.}~\bibnamefont {Shi}}, \bibinfo {author}
  {\bibfnamefont {L.}~\bibnamefont {Patthey}}, \bibinfo {author} {\bibfnamefont
  {G.}~\bibnamefont {Balakrishnan}}, \bibinfo {author} {\bibfnamefont
  {P.}~\bibnamefont {Hofmann}}, \bibinfo {author} {\bibfnamefont
  {R.}~\bibnamefont {Arita}}, \bibinfo {author} {\bibfnamefont
  {N.}~\bibnamefont {Nagaosa}}, \ and\ \bibinfo {author} {\bibfnamefont
  {F.}~\bibnamefont {Baumberger}},\ }\href {\doibase 10.1038/ncomms2162}
  {\bibfield  {journal} {\bibinfo  {journal} {Nat. Commun.}\ }\textbf {\bibinfo
  {volume} {3}},\ \bibinfo {pages} {1159} (\bibinfo {year} {2012})}\BibitemShut
  {NoStop}%
\bibitem [{\citenamefont {Copie}\ \emph {et~al.}(2009)\citenamefont {Copie},
  \citenamefont {Garcia}, \citenamefont {B{\"{o}}defeld}, \citenamefont
  {Carr{\'{e}}t{\'{e}}ro}, \citenamefont {Bibes}, \citenamefont {Herranz},
  \citenamefont {Jacquet}, \citenamefont {Maurice}, \citenamefont {Vinter},
  \citenamefont {Fusil}, \citenamefont {Bouzehouane}, \citenamefont
  {Jaffr{\`{e}}s},\ and\ \citenamefont {Barth{\'{e}}l{\'{e}}my}}]{Copie2009}%
  \BibitemOpen
  \bibfield  {author} {\bibinfo {author} {\bibfnamefont {O.}~\bibnamefont
  {Copie}}, \bibinfo {author} {\bibfnamefont {V.}~\bibnamefont {Garcia}},
  \bibinfo {author} {\bibfnamefont {C.}~\bibnamefont {B{\"{o}}defeld}},
  \bibinfo {author} {\bibfnamefont {C.}~\bibnamefont {Carr{\'{e}}t{\'{e}}ro}},
  \bibinfo {author} {\bibfnamefont {M.}~\bibnamefont {Bibes}}, \bibinfo
  {author} {\bibfnamefont {G.}~\bibnamefont {Herranz}}, \bibinfo {author}
  {\bibfnamefont {E.}~\bibnamefont {Jacquet}}, \bibinfo {author} {\bibfnamefont
  {J.-L.}\ \bibnamefont {Maurice}}, \bibinfo {author} {\bibfnamefont
  {B.}~\bibnamefont {Vinter}}, \bibinfo {author} {\bibfnamefont
  {S.}~\bibnamefont {Fusil}}, \bibinfo {author} {\bibfnamefont
  {K.}~\bibnamefont {Bouzehouane}}, \bibinfo {author} {\bibfnamefont
  {H.}~\bibnamefont {Jaffr{\`{e}}s}}, \ and\ \bibinfo {author} {\bibfnamefont
  {A.}~\bibnamefont {Barth{\'{e}}l{\'{e}}my}},\ }\href {\doibase
  10.1103/PhysRevLett.102.216804} {\bibfield  {journal} {\bibinfo  {journal}
  {Physical Review Letters}\ }\textbf {\bibinfo {volume} {102}},\ \bibinfo
  {pages} {216804} (\bibinfo {year} {2009})}\BibitemShut {NoStop}%
\bibitem [{\citenamefont {Meier}\ \emph {et~al.}(2011)\citenamefont {Meier},
  \citenamefont {Petrov}, \citenamefont {Mirhosseini}, \citenamefont {Patthey},
  \citenamefont {Henk}, \citenamefont {Osterwalder},\ and\ \citenamefont
  {Dil}}]{Meier2011}%
  \BibitemOpen
  \bibfield  {author} {\bibinfo {author} {\bibfnamefont {F.}~\bibnamefont
  {Meier}}, \bibinfo {author} {\bibfnamefont {V.}~\bibnamefont {Petrov}},
  \bibinfo {author} {\bibfnamefont {H.}~\bibnamefont {Mirhosseini}}, \bibinfo
  {author} {\bibfnamefont {L.}~\bibnamefont {Patthey}}, \bibinfo {author}
  {\bibfnamefont {J.}~\bibnamefont {Henk}}, \bibinfo {author} {\bibfnamefont
  {J.}~\bibnamefont {Osterwalder}}, \ and\ \bibinfo {author} {\bibfnamefont
  {J.~H.}\ \bibnamefont {Dil}},\ }\href {\doibase
  10.1088/0953-8984/23/7/072207} {\bibfield  {journal} {\bibinfo  {journal}
  {Journal of Physics: Condensed matter}\ }\textbf {\bibinfo {volume} {23}},\
  \bibinfo {pages} {072207} (\bibinfo {year} {2011})}\BibitemShut {NoStop}%
\bibitem [{\citenamefont {Plumb}\ \emph {et~al.}(2014)\citenamefont {Plumb},
  \citenamefont {Salluzzo}, \citenamefont {Razzoli}, \citenamefont
  {M{\aa}nsson}, \citenamefont {Falub}, \citenamefont {Krempasky},
  \citenamefont {Matt}, \citenamefont {Chang}, \citenamefont {Schulte},
  \citenamefont {Braun}, \citenamefont {Ebert}, \citenamefont {Min{\'{a}}r},
  \citenamefont {Delley}, \citenamefont {Zhou}, \citenamefont {Schmitt},
  \citenamefont {Shi}, \citenamefont {Mesot}, \citenamefont {Patthey},\ and\
  \citenamefont {Radovi{\'{c}}}}]{Plumb2014}%
  \BibitemOpen
  \bibfield  {author} {\bibinfo {author} {\bibfnamefont {N.~C.}\ \bibnamefont
  {Plumb}}, \bibinfo {author} {\bibfnamefont {M.}~\bibnamefont {Salluzzo}},
  \bibinfo {author} {\bibfnamefont {E.}~\bibnamefont {Razzoli}}, \bibinfo
  {author} {\bibfnamefont {M.}~\bibnamefont {M{\aa}nsson}}, \bibinfo {author}
  {\bibfnamefont {M.}~\bibnamefont {Falub}}, \bibinfo {author} {\bibfnamefont
  {J.}~\bibnamefont {Krempasky}}, \bibinfo {author} {\bibfnamefont {C.~E.}\
  \bibnamefont {Matt}}, \bibinfo {author} {\bibfnamefont {J.}~\bibnamefont
  {Chang}}, \bibinfo {author} {\bibfnamefont {M.}~\bibnamefont {Schulte}},
  \bibinfo {author} {\bibfnamefont {J.}~\bibnamefont {Braun}}, \bibinfo
  {author} {\bibfnamefont {H.}~\bibnamefont {Ebert}}, \bibinfo {author}
  {\bibfnamefont {J.}~\bibnamefont {Min{\'{a}}r}}, \bibinfo {author}
  {\bibfnamefont {B.}~\bibnamefont {Delley}}, \bibinfo {author} {\bibfnamefont
  {K.-J.}\ \bibnamefont {Zhou}}, \bibinfo {author} {\bibfnamefont
  {T.}~\bibnamefont {Schmitt}}, \bibinfo {author} {\bibfnamefont
  {M.}~\bibnamefont {Shi}}, \bibinfo {author} {\bibfnamefont {J.}~\bibnamefont
  {Mesot}}, \bibinfo {author} {\bibfnamefont {L.}~\bibnamefont {Patthey}}, \
  and\ \bibinfo {author} {\bibfnamefont {M.}~\bibnamefont {Radovi{\'{c}}}},\
  }\href {\doibase 10.1103/PhysRevLett.113.086801} {\bibfield  {journal}
  {\bibinfo  {journal} {Physical Review Letters}\ }\textbf {\bibinfo {volume}
  {113}},\ \bibinfo {pages} {086801} (\bibinfo {year} {2014})}\BibitemShut
  {NoStop}%
\bibitem [{\citenamefont {Wang}\ \emph {et~al.}(2015)\citenamefont {Wang},
  \citenamefont {McKeown~Walker}, \citenamefont {Tamai}, \citenamefont
  {Ristic}, \citenamefont {Bruno}, \citenamefont {de~la Torre}, \citenamefont
  {Ricco}, \citenamefont {Plumb}, \citenamefont {Shi}, \citenamefont
  {Hlawenka}, \citenamefont {S{\'{a}}nchez-Barriga}, \citenamefont
  {Varykhalov}, \citenamefont {Kim}, \citenamefont {Hoesch}, \citenamefont
  {King}, \citenamefont {Meevasana}, \citenamefont {Diebold}, \citenamefont
  {Mesot}, \citenamefont {Radovic},\ and\ \citenamefont
  {Baumberger}}]{Wang2015}%
  \BibitemOpen
  \bibfield  {author} {\bibinfo {author} {\bibfnamefont {Z.}~\bibnamefont
  {Wang}}, \bibinfo {author} {\bibfnamefont {S.}~\bibnamefont
  {McKeown~Walker}}, \bibinfo {author} {\bibfnamefont {A.}~\bibnamefont
  {Tamai}}, \bibinfo {author} {\bibfnamefont {Z.}~\bibnamefont {Ristic}},
  \bibinfo {author} {\bibfnamefont {F.~Y.}\ \bibnamefont {Bruno}}, \bibinfo
  {author} {\bibfnamefont {A.}~\bibnamefont {de~la Torre}}, \bibinfo {author}
  {\bibfnamefont {S.}~\bibnamefont {Ricco}}, \bibinfo {author} {\bibfnamefont
  {N.~C.}\ \bibnamefont {Plumb}}, \bibinfo {author} {\bibfnamefont
  {M.}~\bibnamefont {Shi}}, \bibinfo {author} {\bibfnamefont {P.}~\bibnamefont
  {Hlawenka}}, \bibinfo {author} {\bibfnamefont {J.}~\bibnamefont
  {S{\'{a}}nchez-Barriga}}, \bibinfo {author} {\bibfnamefont {A.}~\bibnamefont
  {Varykhalov}}, \bibinfo {author} {\bibfnamefont {T.~K.}\ \bibnamefont {Kim}},
  \bibinfo {author} {\bibfnamefont {M.}~\bibnamefont {Hoesch}}, \bibinfo
  {author} {\bibfnamefont {P.~D.~C.}\ \bibnamefont {King}}, \bibinfo {author}
  {\bibfnamefont {W.}~\bibnamefont {Meevasana}}, \bibinfo {author}
  {\bibfnamefont {U.}~\bibnamefont {Diebold}}, \bibinfo {author} {\bibfnamefont
  {J.}~\bibnamefont {Mesot}}, \bibinfo {author} {\bibfnamefont
  {M.}~\bibnamefont {Radovic}}, \ and\ \bibinfo {author} {\bibfnamefont
  {F.}~\bibnamefont {Baumberger}},\ }\href@noop {} {\  (\bibinfo {year}
  {2015})},\ \Eprint {http://arxiv.org/abs/1506.01191v1} {arXiv:1506.01191v1}
  \BibitemShut {NoStop}%
\end{thebibliography}%
\end{document}